# The Ultraviolet Index is well estimated by the terrestrial irradiance at 310nm

**P.D. Kaplan[1] and E.L.P. Dumont[1,2]**

[1]*YouV Labs. Inc ("Shade"), 340 Kingsland St, Nutley, NJ, 07110, USA*
[2]*Hackensack-Meridian Health Center for Discovery and Innovation, Nutley, NJ 07110, USA*
*Correspondence to: peter@shade.io or em@shade.io.*

Ultraviolet (UV) exposure significantly contributes to non-melanoma skin cancer. In the context of health, UV exposure is the product of time and the UV Index (UVI), a weighted sum of the irradiance $I(\lambda)$ over all wavelengths from $\lambda$=250 to 400nm. (1) In our analysis of the United States Environmental Protection Agency's UV-Net database of over four-hundred thousand spectral irradiance measurements taken over several years, we found that the UVI is well estimated by UVI = $77 I_{310}$. (2) To better understand this result, we applied an optical atmospheric model of the terrestrial irradiance spectra and found that it applies across a wide range of conditions.

## 1. Introduction

Non-melanoma skin cancer is highly prevalent despite a strong understanding that its risk is driven by ultraviolet light (UV) exposure.[1,2] As UV is not visible, its damage is wavelength dependent, and skin cancer is significantly delayed from UV exposure, the public's understanding of the links between behavior and risk are poor. Wearable UV monitors have the potential to be a source of data to drive skin-cancer risk reduction. To our knowledge there are, however, no currently available sensors that are both sufficiently accurate and inexpensive for individual use.[5] This is due to the complex wavelength-dependent skin sensitivity to UV which these monitors must replicate in order to be accurate. The relative sensitivity of the skin to each wavelength, also called the erythema action spectrum, is included in the UV Index (UVI), an irradiance unit adopted by the WHO.[3,4] Here we present a novel way to estimate the UV index, by analyzing over 400,000 UV spectra collected over 10 years at multiple sites.[7] We find that a narrow band irradiance detector at 310 nm is sufficient to measure UVI.[6] Using a simple atmospheric model[8] to explore the range of applicability of the result we predict the strategy's success across essentially all terrestrial atmospheric conditions. This feasible, accurate, and wearable UV-monitor design has promise for use in data-driven behavior modification.

    The UVI is calculated (Fig. 1) by adding the erythema-weighted solar irradiance at each wavelength and then dividing the total by a reference value of 25 mW/m². (*6*) By design, the UV index is dependent on the local solar spectrum which in turn is determined by instantaneous cloud cover and other environmental details, features that are ignored when using published values of UVI. (*7*) Additionally, detectors can track the duration, improving accuracy. (*8*, *9*)

    There have been many attempts to develop commercial, hand-held, or wearable low-cost UV dosimeters. (*10*) These detectors, however, have been largely inaccurate (*11*) as there is significant mismatch between their spectral sensitivity and the erythema action spectrum (Fig. 1d). Indeed, because there is such spectral variation, by the latitude, altitude, time of day, season, local weather, and hyper-local environment, it is generally understood that any detector whose spectral sensitivity is not exactly the erythema action spectrum will not be able to accurately predict the UVI. (*12*)

In this report, we present a novel way to accurately measure the UVI that is widely applicable across geographies, seasons, and atmospheric conditions. This work was originally undertaken to produce a minimum technical target for a machine learning design of an optimal low-cost UVI detector. We find that a remarkably accurate detector can be designed using only the irradiance at 310 nm. The result emerges from an analysis of a large database of spectral measurements of sunlight in several US locations across many years and is further validated by an atmospheric model.

## 2. Background and Methods

The UVI is well understood as an integral over three multiplied spectra: sunlight above the atmosphere, loss during transit due to aerosol scattering and ozone absorption, and the erythema action spectrum (Figure 1). The spectrum of solar light impinging on the atmosphere is relatively stable and well documented. Loss of UV in the atmosphere is mostly due to aerosol scattering primarily from water droplets including clouds, and to ozone absorption. The erythema action spectrum weights each wavelength according to its potential to damage skin ([13], [14]). This spectrum has been adopted by the World Health Organization.

Technical designs for UVI detectors have taken two routes. The first, pioneered by Robertson and Berger in 1976, uses a meter that responds to each wavelength according to the erythema action spectrum. ([15]) This approach is both expensive and bulky as it requires careful combinations of filters and detectors. The second, used by most consumer-grade detectors, is based on a UV photodiode and the hope that although spectrally mismatched, it can be accurate enough. The nature of UV spectral fluctuations, which we will discuss in detail below, has defeated this approach. To look for a new approach, we analyzed the US Environmental Protection Agency's UV-Net database. ([16]) This database includes four hundred thousand UV irradiance spectra collected by atmospheric scientists across nine sites between 1996 and 2004, using Brewer spectrometers that collect data in 0.5nm bandwidth channels from 286-363nm. Sites range from near sea level to mountainous elevations, and from urban to rural. UV-Net data have been used for both atmospheric and clinical research. ([16], [17])

For each spectrum, we calculated the UVI and for each wavelength $\lambda$, we performed a linear regression of UVI ~ $I_\lambda$ where $I_\lambda$ is the irradiance. To compare wavelengths, we used the relative difference between the spectrum's UVI and the UVI predicted by the linear fits. This metric was introduced by Correa et al. in the first study comparing the performance of hand-held UV sensors. ([11]) As a measure of accuracy, we calculated the percentage of measurements within a given tolerance (%) of the ground-truth value (Fig. 2). The model most commonly yielding both the lowest relative difference and the highest overall accuracy is UVI = 76.6 $I_{310}$-0.02, 95% confidence intervals (76.62, 76.66) UVI-nm-m$^2$/W and (-0.0166, -0.0145) UVI; $R^2$=0.99.

## 3. Results and Discussion

To illustrate this discovery, we present a simple graphical derivation in Figure 3a, a plot of all data from a single spectrometer over a single day when the UVI exceeded 0.5. The spectra are complex even after erythemal weighting (Fig. 3b). However, when each weighted irradiance spectrum is divided by the UVI, (Fig. 3c) they converge only at 310 nm. The code in the supplementary material enables the reader to render figure 3 using data from any day at any UV-Net site.

To put this performance in the context of the accuracy of other detectors, we model the performance of a detector measuring only irradiance at 310+/- 0.5nm to three low-cost detectors, (SiC) a bare silicon carbide photodiode (SGS01S-18, SGLux), (VEML) a packaged detector sold for measuring UVI (VEML6075, Vishay), and (BB) a broad band optical detector which we model in combination with a flat bandpass filter from 280-400nm (OPT3002, Texas Instruments). The wavelength dependent sensitivity for each device is extracted from its publicly available data sheet. For each detector $d$ we calculated $I_d$, the current expected under each of the UV-Net spectra, built a linear model UVI~$I_d$, and tabulated the accuracy of the model as described above. The 310-detector accuracy at a 10% tolerance $A_{310}^{10}$ exceeds that of non-specialized detectors by a factor of 3: $(A_{310}^{10}, A_{SiC}^{10}, A_{VEML}^{10}, A_{BB}^{10}) = (65, 19, 19, 19)\%$.

To further investigate the origins and limits of the relationship we use a simple, published, single-layer atmospheric model. (*18*) The model irradiance $I_m$ depends on only 3 adjustable parameters $A$, $B$, and $z$:

$$I_m(\lambda) = exp\left(-A + B(\lambda - 320) - z\frac{\sigma(\lambda)}{\cos\phi}\right) I_{solar}(\lambda) \tag{1}$$

The parameters are: $A$ (>0), the aerosol loss, $B$ (nm$^{-1}$) a scattering factor which grossly captures the wavelength dependence of Rayleigh scattering from water droplet size of various sizes, and $z$ (>0, molecules/cm²) the amount of atmospheric ozone. The model uses several known quantities including σ, the ozone cross-section (Fig 1b), an adjustment to the optical path through the ozone layer $1/\cos\varphi$ where $\varphi$ (between 0 and π/2) is the solar zenith angle, and Isolar, the air mass zero (AM0) solar irradiance spectrum above the atmosphere (Fig 1a). To demonstrate the remarkable ability of this model to capture the spectral variation of solar light, in Figure 1c, we show the overlap of this model's predictions on a reference irradiance spectrum. Each of the 405,000 UV-Net spectra were fit to equation 1. The resulting distribution was used to estimate the naturally occurring range of parameters $A$, $B$, and $z$. We then generated a sample of points in the 3-dimensional parameter space $A$, $B$, and $z$ covering a range centered on the observed values and extending along each axis by two standard deviations. In Figure 4 we show the error in the $A$-$z$ projection of this parameter space by coloring each point according to its engineering accuracy, |UVI-77$I_{m310}$|/UVI. Similar projections in the ($A$, $B$), and ($B$, $z$) planes show no additional clear boundaries for the model. This diagram illustrates the expected range of accuracy for the 310nm design. Naturally occurring ozone levels are generally in the range of 250-500 Dobson Units and average 300, the equivalent of a 3mm thick layer of ozone gas at zero degrees Celsius and 1 atmosphere. We see that under ozone holes, defined as ozone thickness less than 220 DU, the model begins to underestimate the UVI. The lowest observed ozone level is 93 DU. Using the solar irradiance at 310 nm to predict the UVI should be valid over a wide range of clear and cloudy skies. Remarkably, the conditions under which the relationship UVI=77$I_{310}$ errs by more than a factor of 1.5 rarely occur on earth.

## 4. Conclusion

Reducing a spectral analysis to one or a few wavelengths is a considerable practical simplification for any application. For example, low-cost blood oxygen measurements depend on the ability to operate with two wavelengths. Here, we have described a novel, surprising, and simple relationship between the irradiance of the sun at 310 nm and the response of human skin to UV. The practical realization of this relationship is straightforward: a narrow-band pass filter centered on 310 nm over a photodiode. We have also described a useful method to evaluate the design performance of UV sensors using their spectral sensitivity and a large database of publicly available spectra. We confirmed that our result is sensible using a simple atmospheric model over a wide range of conditions, providing additional confidence that the relationship will hold over untested weather conditions and geographies.

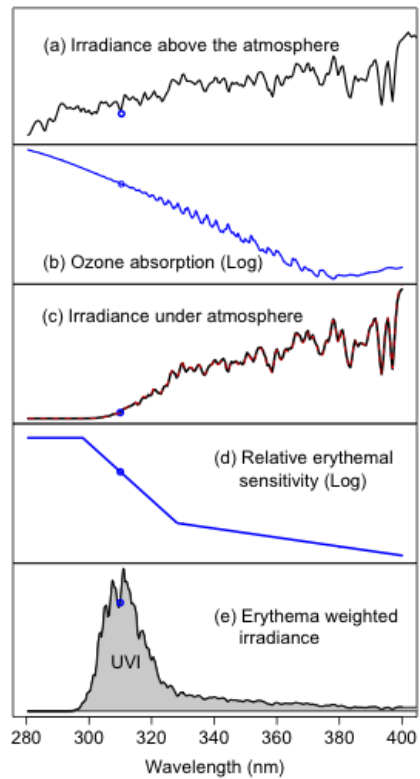

**Fig. 1.** Wavelength dependent contributions to the UV Index. (a) The AM0 solar spectrum spectrum (W/m²/nm), is attenuated in the atmosphere by both aerosol loss (light scattering, not pictured) from water droplets including clouds and by absorption, mostly due to (b) ozone (cross section in cm²/molecule). (c) The resulting AM1.5 solar irradiance on the earth's surface, black line, and a fit mentioned in the discussion, red-dashed line, that quantitatively follows each wiggle in the solar spectrum (d) must be weighted by multiplying by the erythema action spectrum (dimensionless) which is the relative sensitivity of skin to each wavelength. (e) The product is the biologically-relevant erythema-weighted irradiance spectrum. The UVI is given by the shaded area under this curve. The value of each quantity at 310 nm is highlighted (blue dots). (a) and (c) are drawn from ASTM-G159-98, (*19*). The erythema action spectrum is ISO/CIE 17166:2019.

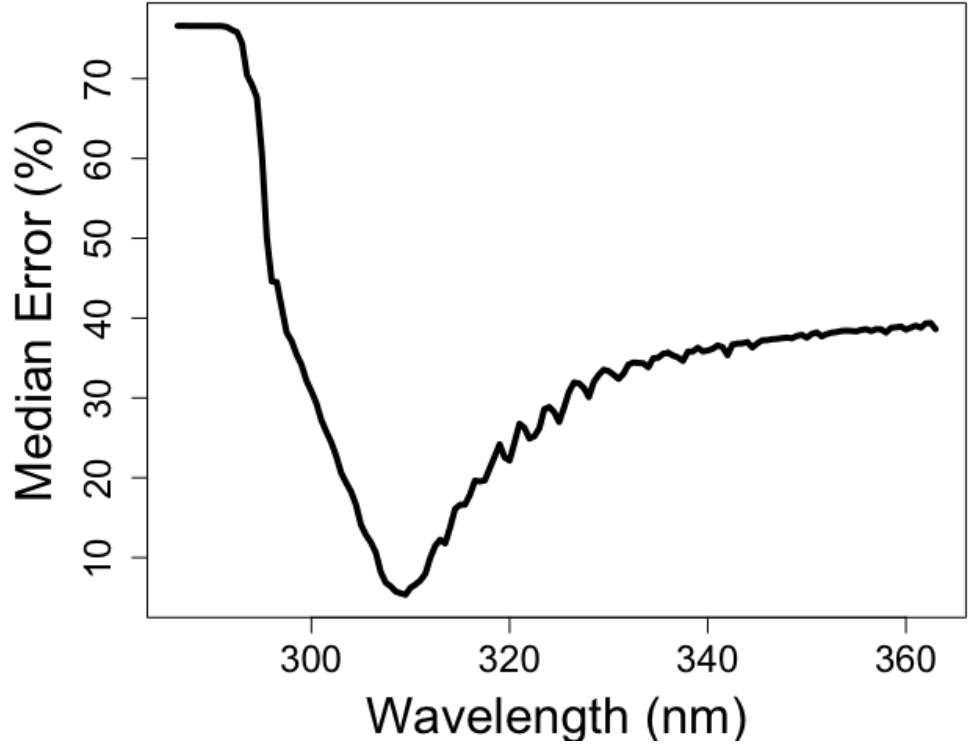

**Fig. 2.** The median percent error in UVI prediction across all 405,000 UV-Net spectra, modeled as an ideally-calibrated linear single-wavelength detector.

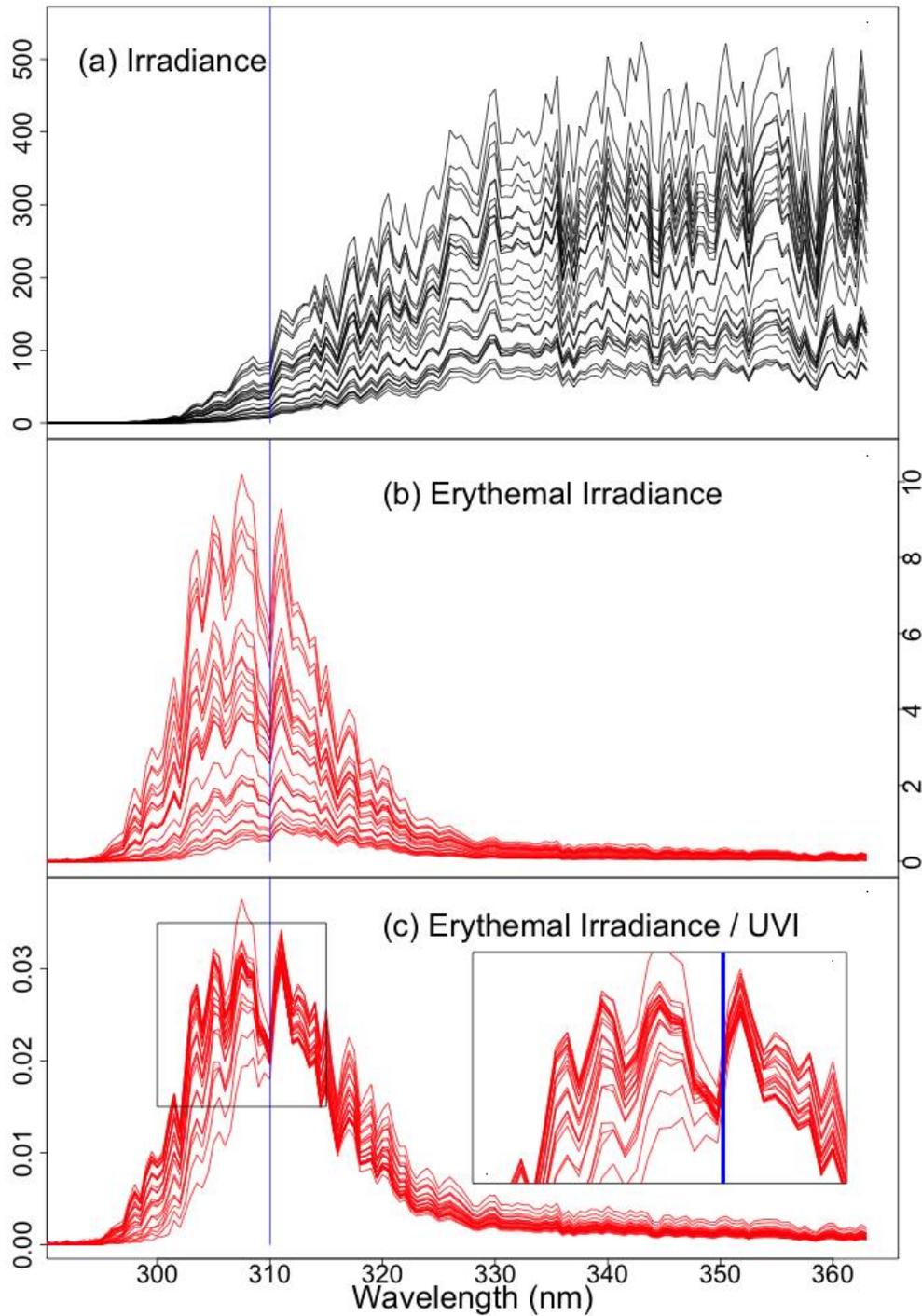

**Fig. 3.** Graphical derivation using (a) all UV spectra (W/m$^2$/nm) from Acadia National Park, June 21, 2000 with UVI>0.5. The spectra vary through the day and do not, in any simple way, resemble the standard AM1.5 solar spectra (Fig 1c). (b) The same spectra weighted by the erythema action spectrum (Fig 1d), (c)These spectra normalized by UVI (W/m$^2$/nm/UVI). After normalization, the spectra collapse, nearly to a point, at 310 nm.

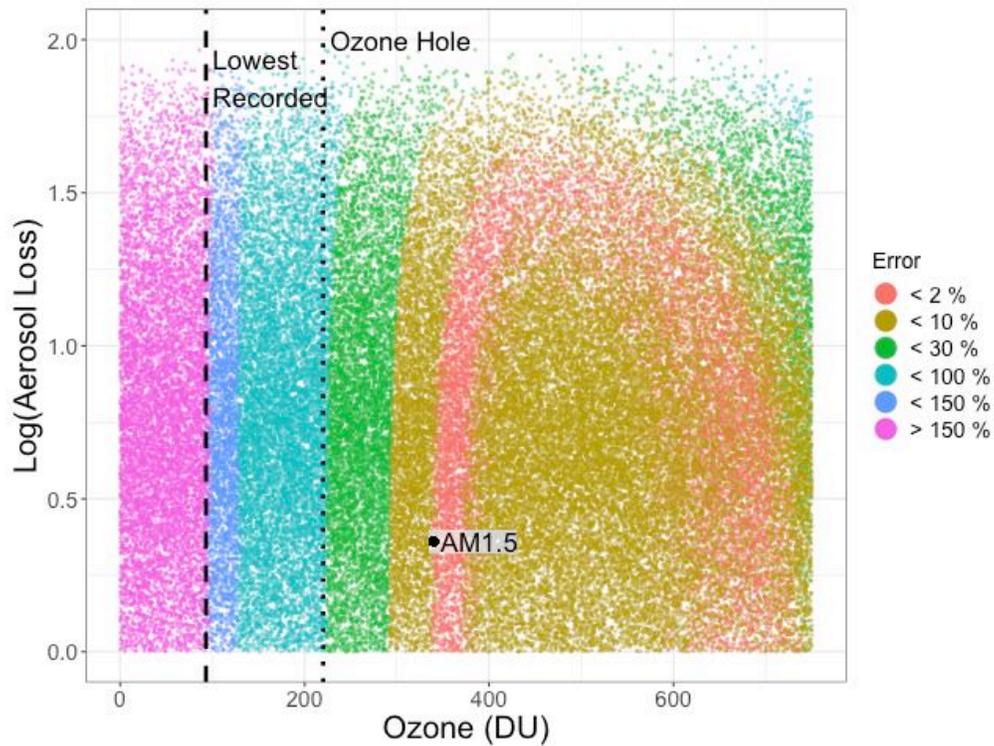

**Fig. 4.** The error predicted by an atmospheric optical model (equation 1) for using the approximation UVI = 76.6 $I_{310}$ nm. For reference we highlight the location of the standard AM1.5 Spectrum (Fig. 1c) (black dot).

**Funding:**

National Science Foundation awards 1746461 and 1951189.

**Acknowldegements**

We are grateful to the UV-NET team for the data behind this research, in particular Patrick Disterhof, contracted to NOAA, for help in obtaining independent validation data.

**Disclosures**

Authors are shareholders in Shade, a wearable UV sensor company.

**Data Availability**

All UV data are from the UV-Net website of the U.S. Environmental Protection Agency and are now available from the Canadian World Ultraviolet and Ozone Data Center.